\documentclass[prc,nofootinbib]{revtex4} 
 \usepackage{epsfig}
 \usepackage{graphicx}
 \usepackage{dcolumn}
 \usepackage{amsfonts}
 \usepackage{amsmath}
 \usepackage{amssymb}
 \usepackage{color}
 \usepackage{mathrsfs}
  \newcommand{\beq}{\begin{equation}}
 \newcommand{\eeq}{\end{equation}}
 \newcommand{\ba}{\begin{array}}
 \newcommand{\ea}{\end{array}}
 \newcommand{\beqa}{\begin{eqnarray}}
 \newcommand{\eeqa}{\end{eqnarray}}
 \newcommand{\bal}{\begin{align}}

 \def\Rb{\mathbb{R}}

\usepackage{docs}%
\usepackage{bm}%
\expandafter\ifx\csname package@font\endcsname\relax\else
 \expandafter\expandafter
 \expandafter\usepackage
 \expandafter\expandafter
 \expandafter{\csname package@font\endcsname}%
\fi

\begin{document}

\title{Nuclear shape coexistence from the perspective of an algebraic many-nucleon version of the Bohr-Mottelson unified model}
\author{David J.~Rowe}
\affiliation{Department of Physics, University of Toronto, Toronto, ON M5S 1A7, Canada}
\date{September, 2019}

\begin{abstract}
%
	A fully quantal algebraic version of the Bohr-Mottelson unified model is presented with the important property that its quantisation is defined by its irreducible unitary representations which span the many-nucleon Hilbert space of every nucleus.
	The model is uniquely defined by the requirement that its Lie algebra of observables includes the nuclear quadrupole moments and kinetic energy.
	It then follows that there can be no non-zero isoscalar E2 transitions between any states belonging to its different irreducible representations and, as a result, the states of the model are uniquely defined with the property that observed transitions between rotational states of nuclei are to be expressed in terms of mixtures of the model irreps.
	The algebraic version of the unified model parallels the Bohr-Mottelson model in most respects, including the possibility of including the effects of Coriolis and centrifugal forces as subsequent perturbations.
	However, it corrects its treatment of angular momentum quantisation and no longer uses an over-complete set of coordinates.
	These changes have significant implications for the dynamics of nuclear rotations which are hidden when its moments of inertia are considered as 
inertial masses in the standard expression of rotational kinetic energies.
	Thus, the developments put a new perspective on the phenomenon of shape coexistence.

\end{abstract}	

\maketitle
{\bf PACS numbers:} 21.60.Ev,  21.60.Fw, 21.60.Cs,  24.10.Cn.   

{\bf Keywords:} Unified model. Quantum mechanics of rotations. Algebraic mean-field theory.  Nuclear shape coexistence.\\

\section{Introduction}
Nuclei are of special interest among the many-particle systems that have contributed to the development of quantum mechanics.	
	They have shell structures, similar to those of atoms, and rotational states with properties  between those of molecules and superfluids.
	Such properties and, in particular, the prevalence of deformed nuclei and rotational structures throughout the nuclear periodic table \cite{BohrM75, HeydeIWWM83, WoodHNHvD92, HeydeW11, WoodH16eds} stimulate attempts to understand the dynamics of nuclear rotations and motivate development of the many-nucleon quantum mechanics of rotational states in deformed nuclei.

It is understood that there are small but non-zero perturbations in the classical mechanics of a finite system due to centrifugal and Coriolis forces in a slowly rotating frame of reference.
	In accord with Born-Oppenheimer theory \cite{BornO27x}, it is also understood that these inertial forces are likewise minor perturbations of the low angular-momentum rotational states of nuclei in quantum mechanics.
	Thus, Bohr and Mottelson \cite{BohrM53c} introduced their highly influential unified model of nuclear rotations and shape vibrations characterised by a non-spherical intrinsic state with relatively high-energy vibrational excitations and low-energy rotational states with wave functions expressed as functions of the orientation angles of the intrinsic state and with moments of inertia treated as adjustable parameters.
	The coupling of the low-energy rotational and high-energy vibrational states were  considered, in this model, to be minor perturbations as they would be, in an adiabatic limit, in classical mechanics.
	This is appropriate.
	However, there are fundamental differences between translations and rotations in quantum and classical mechanics, due to the quantisation of angular momentum in quantum mechanics, which are ignored in the unified model. 
	A fundamental difference is that, whereas the wave functions of a nucleus can be expressed as products of centre-of-mass and intrinsic wave functions, they can only be expressed as products of rotational and intrinsic wave functions, as assumed in the unified model, in  a classical limit of quantum mechanics.

A related concern, is the presumption that nuclear rotational energies are kinetic energies, as implied  \cite{BohrM55} in the interpretation of the moments of inertia of the Inglis cranking model \cite{Inglis54, Inglis55} as being linear combinations of those for rigid-body and irrotational flow rotations.
	More sophisticated cranking models \cite{ThoulessV62} and models with the inclusion of superconducting pairing interactions \cite{Belyaev61} were similarly interpreted and
the moments of inertia of essentially all rotational nuclei have been observed, in support of this cranking model interpretation \cite{NathanN65}, to lie between those of irrotational superfluid rotations and the rigid-body flows of a viscous fluid.

The dynamics of nuclear rotations is expected to be model dependent.
	However, as this paper shows, it is possible to define an optimal algebraic many-nucleon version of the unified model with the property that there are no isoscalar E2 transitions between the states of its different irreps (irreducible representations).
	Observed rotational states are then mixtures of its irreps.
	A useful result, shown in section III, is that this optimal model is the already well-known  symplectic model which now acquires an enhanced significance.
	
Early symplectic model calculations were made
\cite{RosensteelR77a, RosensteelR80, DraayerWR84, RosensteelDW84, ParkCVRR84}
to derive the properties of nuclear rotational states with model interactions in truncated finite-dimensional shell-model spaces.	
	In particular, fits were obtained to both the energy levels and E2 transitions of $^{20}$Ne \cite{RosensteelR77a} and four rare-earth nuclei \cite{ParkCVRR84}, without the use of effective charges, and it was determined that the so-called rotational energies of the  nuclei in these calculations were mixtures of kinetic and potential energies of comparable magnitude.
	Similar results were later obtained for $^{166}$Er by Bahri \cite{BahriR00}.
	However, while significant and suggestive, these results with model Hamiltonians in truncated spaces could not be considered definitive in all respects.
	Recall, for example, that the SU(3) model \cite{Elliott58ab}, which is equivalent to a symplectic model with a simple Hamiltonian restricted to the highly truncated shell model space of a single spherical harmonic-oscillator shell, gives fits to observed rotational states in light nuclei with excitation energies that are 100$\%$ potential energies.
	Thus, the interactions and nucleon charges of the SU(3) model are considered as effective interactions and effective charges, respectively, and its rotor-like states can be understood as the projected images of a realistic rotor model onto the many-nucleon space of a single spherical harmonic-oscillator shell.
	Similarly, the results of the calculations by Park, Bahri et al.\  
\cite{ParkCVRR84, BahriR00}, while suggestive, can be attributed to the use of an effective interaction in a much larger but still truncated shell-model space.

\section{The dynamics of the Bohr-Mottelson unified model}	
%
In the Bohr-Mottelson unified model \cite{BohrM53c, BohrM75} the rotations and shape vibrations of a nucleus are characterised by an intrinsic state that corresponds to a classical limit of a quantum rotor model at rest.
	Thus, the unified model is a hybrid of classical and quantal mechanics.
	Its intrinsic state has excited vibrational states corresponding to those of its quantised normal-mode vibrations and its rotational states are described by multiplying these states by rotational wave functions, which are functions of the orientation angles of the intrinsic ground state.
	All the states of the ground state rotational band of the unified model are thereby assigned a common intrinsic state and, hence, common potential energies.
	The remarkable successes of the unified model, with moments of inertia adjusted to fit the data, nevertheless resulted in its widespread acceptance.

Unfortunately, there is negligible experimental information on the nature of nuclear rotations.
	There was early optimism that the current flows in rotating nuclei could be determined from transverse electron scattering cross sections \cite{KunzM79, Radomski76, FlecknerKMW80, MoyaDeGuerra80, BerdichevskySMdGNS88, MikhailovBQ96}.
	However, because of the dominance of much larger longitudinal cross sections, the available cross sections have proved to be too imprecise to be of much value.
	
In seeking to understand the dynamics of nuclear rotations
numerous investigations were pursued, over a period of many years, to identify a maximal separation of the nuclear kinetic energy into intrinsic and rotational components in the hope of understanding the 
unified model's successes and the observed values of nuclear moments of inertia in the adiabatic limits of negligible centrifugal and Coriolis forces.
	Complete decompositions of the nuclear kinetic energy operator were ultimately obtained in terms of collective coordinates and  momenta \cite{Zickendraht71, FilippovOS72, GulshaniR76, WeaverCB76}, and summarised in reference \cite{RoweR79}.  
	They did not identify the separation assumed in the unified model, other than in the classical limit of rigid-body rotations, but were successful in pointing the way to a subsequent development of an optimal algebraic version of the unified model, given by the so-called symplectic model \cite{RosensteelR77a, RosensteelR80}, in which the whole many-nucleon kinetic energy is included in its Lie algebra.
	 This was significant because, while the symplectic model admits the possibility that the kinetic energy is a dominant component of nuclear rotational energies, it does not exclude the possibility of major potential energy contributions.
	A particularly important advantage to having the kinetic energy as an element of the symplectic model's Lie algebra is the fact that its presence facilitates the determination of the kinetic energy components of the model's rotational energies with the implication that the remaining energies are potential energies.

\section{An optimal symplectic model of the rotations and shape vibrations of deformed nuclei}
\label{sect:optimal}
The collective dynamics of nuclear rotations and shape vibrations is model dependent,
	However, it is most usefully defined by an algebraic model with a dynamical group having the following  properties:
(i) it has a set of irreps that together span the fully anti-symmetric many-nucleon Hilbert space of a nucleus;
(ii) its Lie algebra is  a sub-algebra of that of the group of all one-body unitary transformations of a many-nucleon Hilbert space; and
(iii) its Lie algebra contains among its elements the many-nucleon kinetic energy, the angular momentum operators, and the monopole/quadrupole moment operators of nuclei. 
	Conditions (i) and (ii) are needed to restrict consideration to legitimate many-nucleon representations.
	In particular, condition (ii) means that its irreps are restricted to those which are fully anti-symmetric as required of a many-fermion model.
	Condition (iii) ensures that the decomposition of the Hilbert space of a nucleus into irreducible collective model subspaces, has the unique and invaluable property that there can be no non-zero matrix elements of any of the operators in  its Lie algebra between any pair of states belonging to different optimal model irreps.
	While the rotational states of physical nuclei are not expected to satisfy condition (iii), its inclusion enables them to be expressed as mixtures of the optimal collective model states to explain their inter-band E2 transitions.
	The properties of such an optimal rotor model are then of primary interest.

A fortuitous result is that the simplest algebraic model with these properties is already known.
	It is the so-called symplectic model \cite{RosensteelR77a, RosensteelR80, Rowe85} which is an algebraic model with an 
${\rm Sp}(3,\Rb) \times {\rm SU(2)}_S\times{\rm SU(2)}_T$ dynamical group in which Sp$(3,\Rb)$ is the symplectic group with Lie algebra spanned, for an $A$-nucleon nucleus, by the operators
\bal
& \hat Q_{ij} = \sum_{n=1}^A \hat x_{ni} \hat x_{nj} , \quad 
\hat P_{ij} =  \sum_{n=1}^A ( \hat x_{ni} \hat p_{nj} + \hat p_{ni} \hat x_{nj}), \label{eq:4.1}  \\
&\hbar \hat L_{ij} =  \sum_{n=1}^A \big(\hat x_{ni}\hat p_{nj}  -\hat x_{nj}\hat p_{ni}\big), 
\quad \hat K_{ij}= \sum_{n=1}^A  \hat p_{ni}  \hat p_{nj} , \label{eq:4.2} 
\end{align}
where $\hat x_{ni}= x_{ni}$ and $\hat p_{ni}= -{\rm i} \hbar \partial /\partial x_{ni}$ with 
$i,j = 1,2,3$, are the position and momentum observables of the $n=1,\dots,A$ nucleons of a nucleus.
	The spin-isospin group ${\rm SU(2)}_S\times{\rm SU(2)}_T$ is included to take account of the neutron and proton spins and  ensure that the combined space, spin and isospin states of an irrep satisfy the anti-symmetry requirements of a many-nucleon nucleus.
	The  $\hat P_{ij}$ operators, which are recognised as infinitesimal generators of shape deformations, were also not among the required elements of the Lie algebra being sought.
	Their inclusion as infinitesimal generators  of the model's dynamical group is nevertheless seen to be appropriate.
	Together, the totally antisymmetric irreps of this algebra are completely labelled by their quantum numbers and together span the Hilbert spaces of nuclei.
	Another invaluable property, shown in the following section, is that the Bohr-Mottelson unified model emerges as a mean-field approximation to the algebraic model with the direct product ${\rm Sp}(3,\Rb) \times {\rm SU(2)}_S\times{\rm SU(2)}_T$ dynamical group.
	
In practical calculations, particularly for light nuclei, corrections should be made for the spurious centre-of-mass degrees of freedom of nuclear wave functions.
	Fortunately, this is simply achieved in an ${\rm Sp}(3,\Rb)\supset {\rm U(3)}$ basis by use of a strategy devised by Gloeckner and Lawson \cite{GloecknerL74} of simply adding a term to the nuclear Hamiltonian so that  the unwanted centre-of-mass excited states end up at an energy well above that of the states of interest.
	

Because the Sp$(3,\Rb)$ Lie group is semi-simple, all its unitary irreps have lowest-weight states.
	Using this property, its unitary irreps can be systematically constructed \cite{Rowe84, RoweRC84, Quesne86, Hecht87}.
	A single nucleon has two elementary representations with states given, respectively, by those of the nucleon in positive and negative parity states of a spherical harmonic oscillator.
	For a many-nucleon nucleus, the 
${\rm Sp}(3,\Rb) \times {\rm SU(2)}_S\times{\rm SU(2)}_T$ irreps are defined on subspaces of  many-nucleon harmonic-oscillator shell models.

The Lie algebra of the symplectic model is defined a follows.
	First express the position and momentum coordinates of the nucleons in terms of 
harmonic-oscillator raising and lowering operators, $\{ c^\dag_{ni}, c_{ni}\}$, by the standard expressions
\beq \hat x_{ni} = \frac{1}{\sqrt{2}\,a} (c^\dagger_{ni}+ c_{ni} ) , \quad
\hat p_{ni} = {\rm i} \hbar\frac{a}{\sqrt{2}} (c^\dagger_{ni} - c_{ni} ) ,
\label{eq:7,xpccdag} \eeq
where $a = \sqrt{M\omega_0/\hbar}$ is a harmonic-oscillator unit of inverse length.
This gives 
\bal 
&  \hat Q_{ij} =
\frac{1}{2a^2}\big(2\hat{\mathcal{Q}}_{ij}  + \hat {\mathcal{A}}_{ij} + \hat {\mathcal{B}}_{ij} \big), \quad \hat P_{ij} 
= {\rm i}\hbar (\hat {\mathcal{A}}_{ij} - \hat {\mathcal{B}}_{ij}) , \\
& \hat K_{ij} = \frac12 a^2 \hbar^2 (2 \hat{\mathcal{Q}}_{ij} 
- \hat {\mathcal{A}}_{ij} - \hat {\mathcal{B}}_{ij} ) , \quad
\hat L_{ij}  = -{\rm i} ( \hat {\mathcal{C}}_{ij} - \hat {\mathcal{C}}_{ji} ) , 
\end{align}
with
\bal
 &\hat {\mathcal{A}}_{ij} = \hat {\mathcal{A}}_{ji} = \sum_{n=1}^A c^\dagger_{ni} c^\dagger_{nj} ,
 \quad \hat {\mathcal{B}}_{ij} = \hat{\mathcal{B}}_{ji} =\sum_{n=1}^A c_{ni} c_{nj} , 
  \label{eq:4.ABops}\\
&\hat {\mathcal{C}}_{ij} =  \sum_{n=1}^A \big( c^\dagger_{ni} c_{nj} 
+ \textstyle\frac12 \delta_{i,j}\big), \quad
\hat{\mathcal{Q}}_{ij} = \textstyle\frac12\big( \hat {\mathcal{C}}_{ij} + \hat {\mathcal{C}}_{ji} \big) .
  \label{eq:4.CQops}
\end{align}
	Thus, it is apparent that the states of the optimal model are defined on sub-spaces of many nucleons in the positive or negative parity states of a three-dimensional harmonic oscillator. 


\section{A classical mean-field perspective}	\label{sect:MFP}
It has long been understood that classical mechanics can be realised as constrained quantum mechanics in which, for example, the classical states of a finite fermion system are restricted to a sub-manifold of quantum mechanical states, known in physics as coherent states. 
	Coherent states were introduced for harmonic-oscillator states by Glauber \cite{Glauber66} and subsequently defined generally by Perelomov and Klauder \cite{Perelomov72, KlauderB-S85}, for any algebraic model with a dynamical group and irreps with lowest-weight states, as the states generated by the transformations of a lowest-weight state of the irrep by group elements. 
 	Such a set of coherent states is described in mathematics as a co-adjoint orbit \cite{Kostant70, Souriau70, Rosensteel86}.
	
The classical manifold of coherent-states of HF (Hartree-Fock) theory, is the set of states generated by the group of all one-body unitary transformations of a lowest-weight independent-particle state.
	It has the useful property of being the set of all independent-particle states.
	The classical Hamiltonian on this manifold is then defined by the expectation values of the quantum mechanical Hamiltonian in its coherent states and the classical Hamiltonian equations of motion, defined by time-dependent HF theory, are identical to the corresponding equations of motion of constrained quantum mechanics as shown in references \cite{RoweB76, RoweR80, RoweR80a} and reviewed in chapter 6 of \cite{RoweWood10}.
	Thus, a classical equilibrium state on this manifold is a state for which the expectation value of the quantum mechanical Hamiltonian is a (local) minimum.	
	Also the classical normal-mode vibrations of a nucleus about its equilibrium state, given by the time-dependent HF equations of motion \cite{Rowe66a, Rowe66b}, are identical to those of the quantum-mechanical random-phase approximation (RPA) of  Bohm and Pines \cite{BohmP53};.
	This becomes apparent when the RPA is expressed in the double-commutator equations-of-motion formalism \cite{Rowe68, Rowebook70}, which is the form in which it is now commonly used in nuclear physics \cite{RingS80}.

The embedding of classical mechanics in quantum mechanics by mean-field methods is  insightful for understanding the physics of quantum systems from a classical perspective; e.g., for exploring the topography of the classical potential energy surface, as for a landscape, in the neighbourhood of its lowest-energy equilibrium state.
	This was initiated by Rowe and Basserman \cite{RoweB74, RoweB76} and Marumori \cite{Marumori77}, who studied 
the valley floor of its classical potential energy surface and by Reinhard and Goeke
\cite{ReinhardG78}, who studied its fall lines.
	It was then recognised \cite{RoweR82, Rowe82}  that, in proceeding upwards along the valley floor, starting from its lowest point, a high point of the valley will be reached following which the valley path begins a descent to another low-energy point.
	It follows that the HF minimisation procedure could also converge on such  local minimal-energy state, which would be orthogonal to the lowest-energy state, but to which the iterative HF procedure could only converge from points in its neighbourhood.
	Such a possibility has been considered by Matsuyanagi and colleagues \cite{MatsuyanagiMNYHS16}.
	The significance of local minimum-energy HF states has not, as yet, been explored.
A promising interpretation,  relates to the many minima given by the  algebraic many-nucleon version of the unified model, given in section \ref{sect:optimal} and further pursued in the following section.	
	
%
When the Lie algebra of all one-body operators is restricted to its 
${\rm Sp}(3,\Rb) \times {\rm SU(2)}_S\times{\rm SU(2)}_T$ symplectic model sub algebra, the classical potential energy HF surface separates into a sum of disconnected energy surfaces, each of which is defined for an irrep of the symplectic model sub algebra.
	It would then not be surprising to find that, if an iterative HF calculation were initiated from the minimum energy state of one these symplectic model irreps,  it converged to a local minimum energy state of essentially the same general form.
	For example, if a HF calculation for $^{16}$O were initiated from a closed-shell state of spherical harmonic oscillator states, it would be expected to converge to a closed-shell state of single-nucleon states with  modified radial wave functions.
	Similarly, if it were initiated from a deformed 4-particle - 4-hole symplectic model minimum-energy state, it would likewise be expected to converge to a modified but still recognisable  4-particle - 4-hole state relative to the closed-shell state.

\section{An algebraic many-nucleon (AMN) unified model} \label{sect:V.SpST}
This section presents a fully quantal  construction of the representations of an AMN unified model as an optimal many-nucleon model within the space of a fully anti-symmetric irrep of an
${\rm Sp}(3,\Rb)\times{\rm SU(2)}_S\times{\rm SU(2)}_T$ dynamical group as defined in section \ref{sect:optimal}.
	The important property of these 
irreps is that, for minimal values of spin and isospin, $S$ and $T$, they are of maximal space symmetry.
	As a result, they are maximally deformed and are expected to be lowered most from their spherical shell-model energies relative to those of less space symmetry \cite{RoweW18}.
	This provides an understanding of the shape coexistence of strongly deformed states among those which, in a spherical shell model, would have lower energies and be less deformed  \cite{HeydeW11,WoodH16eds}.

A lowest-weight state $|\sigma, \omega\rangle$ for a fully anti-symmetric 
${\rm Sp}(3,\Rb) \times {\rm SU(2)}_S\times{\rm SU(2)}_T$ irrep of a nucleus with an Sp$(3,\Rb)$ lowest weight $(\sigma_1,\sigma_2,\sigma_3)$, with 
$\sigma_1\geq \sigma_2 \geq \sigma_3$,  spin $S=S_0$, and isospin $T=T_0$, is defined by the equations
\beqa 
&\hat {\mathcal{C}}_{ij}|\sigma, \omega\rangle =& 0, \quad \text{for $i < j$} ,\\
& \hat {\mathcal{C}}_{ii}|\sigma, \omega\rangle =&\sigma_i |\sigma, \omega\rangle ,
\quad \text{for $i =1,2,3$}.
\eeqa
	It follows that it is the fully anti-symmetric ground state $|\sigma, \omega\rangle$ of a mean-field harmonic-oscillator Hamiltonian $\hat{\mathcal{H}}(\omega)$
for which
\beq \label{eq:4.THO}  
\hat{\mathcal{H}}(\omega)|\sigma, \omega\rangle
 \equiv  \sum^3_{i=1} \hbar\omega_i \hat {\mathcal{C}}_{ii}|\sigma, \omega\rangle
= \sum^3_{i=1} \hbar\omega_i \sigma_i |\sigma, \omega\rangle .
\eeq

The lowest-energy of an Sp$(3,\Rb)$ irrep of  lowest-weight state 
$\sigma = (\sigma_1,\sigma_2,\sigma_3)$ for a given nuclear Hamiltonian $\hat H$ is given by the values of $(\omega_1,\omega_2,\omega_3)$, for which the energy
$\langle \sigma,\omega |\hat H |\sigma,\omega\rangle$ is minimised, and can be determined by Hartree-Fock methods.
	The self-consistency property of HF theory then implies that the values of 
$(\omega_1,\omega_2,\omega_3)$ are given to a high degree of accuracy by the condition that the potential-energy component 
\beq V(x) = \frac12 M \sum_n \left(\omega_1^2 x_{n1}^2 + \omega_2^2 x_{n2}^2 + \omega_3^2 x_{n3}^2\right)  , \label{eq:4.V(x)}\eeq
of the mean-field Hamiltonian $\hat{\mathcal{H}}(\omega)$, of which the state
$|\sigma, \omega\rangle$ is an eigenstate, has maximal overlap with  the density of the minimum-energy state $|\sigma, \omega\rangle$.

	
Given that the surfaces of constant  $V(x)$ potential energy are ellipsoidal and, that the mean values of  $\sum_n x_{ni}^2$ are given by	
\beq
 \langle x_i^2\rangle_{\omega(\sigma)} \equiv \langle \sigma, \omega | \sum_n x_{ni}^2 |\sigma, \omega\rangle  = \frac{\hbar \sigma_i}{M\omega_i}\, , \quad i=1,2,3, \eeq
it follows that a close approximation to
%
%
%
an ellipsoidal equi-density surface for this minimal energy state, is defined by the equation
\beq \frac{x_1^2}{\langle x_1^2\rangle_{\omega(\sigma)}}
+ \frac{x_2^2}{\langle x_2^2\rangle_{\omega(\sigma)}}
+ \frac{x_3^2}{\langle x_3^2\rangle_{\omega(\sigma)}} = {\rm const.}
\eeq
i.e., by 
\beq \frac{\omega_1x_1^2}{\sigma_1} + \frac{\omega_2x_2^2}{\sigma_2} 
 +\frac{\omega_3x_3^2}{\sigma_3}= {\rm const.} \label{eq:equidensity}\eeq
	An equipotential surface for the potential (\ref{eq:4.V(x)}) is  given by
\beq \omega_1^2 x_1^2 + \omega_2^2 x_2^2 +\omega_3^2 x_3^2 
= {\rm const}. \label{eq:equipot}\eeq
Thus, for these surfaces  to have the same ellipsoidal shape, it is required that 
\beq \sigma_1\omega_1 = \sigma_2\omega_2 =\sigma_3\omega_3 . \label{eq:SCcondition} \eeq
	Such a self-consistency relationship has been used for other purposes; e.g., by Bohr and Mottelson \cite{BohrM55, BohrM75}  and Castel et al.\ \cite{CastelRZ90}.
	
When $\sigma_1= \sigma_2 =\sigma_3$ and $S=0$, the lowest-weight state of the symplectic model is  rotationally invariant and in a state of zero orbital angular momentum.
	It is then a many-nucleon $L=S=J=0$, $T=T_0$ ground state of a spherical harmonic oscillator and has 
one-phonon monopole and quadrupole vibrational excitations given by the random phase approximation.

When $\sigma_1 > \sigma_2$ and/or $\sigma_2>\sigma_3$, 
and $S=S_0$ is not necessarily zero, 
the shape-consistent lowest-energy lowest-weight state for an irrep
of the ${\rm Sp}(3,\Rb)\times{\rm SU(2)}_S\times{\rm SU(2)}_T$ dynamical group
has an interpretation as the intrinsic state of an algebraic many-nucleon version of the Bohr-Mottelson unified model.
	Following the Nambu-Goldstone interpretation \cite{NambuJL61, GoldstoneSW62},
such a broken-symmetry (non-rotationally invariant) minimum-energy state is observed to be one state of a multi-dimensional vector space of equal-energy states generated by the rotations of  the broken symmetry state.
	With appropriate development of the required computational techniques, the Hamiltonian for the nucleus could then be diagonalised in a basis for this vector space to give the rotational states of the required algebraic many-nucleon version of the unified model.
	The construction of the rotational states of such a unified model corresponds closely to that of the angular-momentum projection methods of standard HF theory as outlined, for example, by Lee and Cusson \cite{LeeC72, CussonL73}.
	The intrinsic beta and gamma vibrational excited states of this model are also defined  by standard time-dependent mean-field, or equivalent RPA methods.
	Thus, a non-rotationally invariant HF minimum-energy state acquires a natural interpretation as the intrinsic state of a many-nucleon version of the Bohr-Mottelson unified  model with the property that it avoids the introduction of spurious rotational coordinates.

\section{An energy ordering of 
symplectic model irreps and shape coexistence}
In the standard shell model, an energy-ordered basis of independent-particle states is defined by the summed energies of the neutrons and protons in a  spherical harmonic-oscillator mean field with spin-orbit interactions and an added angular-momentum dependent term.
	The ordering is then used in the selection of an active valence space for shell-model calculations of nuclear states.
	This is appropriate for doubly closed-shell nuclei and for states of singly closed-shell nuclei with ground states that give no evidence of being deformed or of belonging to rotational sequences of states. 
	However, even for spherical nuclei,  there are frequent occurrences of strongly deformed states at low-energy excitation energies.
	A textbook example of this is the first excited state of $^{16}$O which is understood to be the $J=0$ ground state of a band of a strongly deformed rotational states with an intrinsic state given by a 4 particle - 4 hole excitation of the spherical $^{16}$O closed-shell state \cite{Morinaga56, BrownG66}.

Consistent with the Nilsson model \cite{Nilsson55, Lamm69} and the widespread observation of nuclear shape coexistence \cite{HeydeW11, WoodH16eds}, HF calculations \cite{BenderHR03, ErlerKR11} imply that most nuclei should have deformed  low-energy states.
	However, a challenge remains to understand observed rotational states of nuclei in terms of interacting many-nucleon quantum mechanics. 
	A meaningful approach is to identify the AMN unified model irreps that best describe the observed properties of a given rotational nucleus and estimate the energies at  which these irreps are expected to be observed.
%
	A useful start is to list the possibilities.
	
Table \ref{tab:order.reps} gives a list of positive parity $S=0$, $T=\tfrac12 (N-Z)$ AMN unified model irreps for three nuclei
ordered by increasing values of the energies $E_\sigma$
of the lowest-energy lowest-weight  states as determined by the self-consistency relationships 
\beq \label{eq:LWtHOstate}
 E_\sigma \hbar\omega_0
 =\langle \sigma,\omega |\hat{\mathcal H}(\omega) | \sigma,\omega\rangle 
= \sum_i \hbar \omega_i \sigma_i  
= 3 (\sigma_1\sigma_2\sigma_3)^\frac13 \hbar \omega_0 , 
\quad \text{with $\omega_0^3= \omega_1\omega_2\omega_3$},
\eeq
for a range of $N_0=\sigma_1+\sigma_2+\sigma_3$ values; these are the states  of maximal space symmetry with a value of $\omega_0$ chosen such that the lowest-weight state has the expected volume for the nucleus under consideration.
\begin{table}[tbh]  
\caption{\label{tab:order.reps} \footnotesize Comparison of  the minimum mean-field energies $E_\sigma=\langle \sigma,\omega |\hat{\mathcal H} | \sigma,\omega\rangle/\hbar\omega_0$,
 given by equation (\ref{eq:LWtHOstate}), and the corresponding value of 
 $\lambda = \sigma_1-\sigma_2$ and $\mu =  \sigma_2-\sigma_3$,
for a range of values of $N_0 = \sigma_1+\sigma_2+\sigma_3$ increasing from the minimum value allowed by the Pauli exclusion principle for states of spin $S=0$ and isospin $T=T_0$.
	The contribution of the centre-of-mass to the energies shown has been removed.
	The representations are restricted to those of positive parity and the
 rows are ordered by increasing values of 
 $E_\sigma=3 (\sigma_1\sigma_2\sigma_3)^\frac13$.
 \vspace{0.2cm} }
  \begin{minipage}[t]{1.8in}
\centerline{${}^{12}${\rm C} }\vspace{0.1cm}
$\begin{array}{|c|c|c|c|c|c|}\hline
{N_0} &  \lambda &  \mu & 2\lambda+ \mu &  E_\sigma   \\ \hline
24.5 &   0 & 4  &  4  & 23.75 \\ 
28.5 & 12 & 0  & 24 & 24.27 \\ 
26.5 &   6 & 2  & 14 & 24.68 \\ 
30.5 & 10 & 2  & 22 & 26.91 \\
32.5 & 12 & 2  & 26 & 27.90 \\
\hline
\end{array}$
\end{minipage}   
\begin{minipage}[t]{1.9in}
\centerline{${}^{16}${\rm O} }\vspace{0.1cm}
$\begin{array}{|c|c|c|c|c|}\hline
{N_0} &  \lambda &  \mu & 2\lambda+ \mu  & E_\sigma   \\ \hline
34.5 &    0 &  0  & 0   & 34.50 \\ 
38.5 &    8 &  4  & 20 & 35.68  \\ 
36.5 &    4 &  2  & 10 & 35.78 \\ 
46.5 &  24 &  0  & 48 & 36.30 \\
42.5 &  16 &  2  & 34 & 36.61\\ 
40.5 &  10 &  4  & 28 & 36.86 \\ 
\hline
\end{array}$\\
\end{minipage} 
\begin{minipage}[t]{1.9in}
\centerline{${}^{168}${\rm Er} }\vspace{0.1cm}
$\begin{array}{|c|c|c|c|c|}\hline
{N_0} &  \lambda &  \mu & 2\lambda+\mu &  E_\sigma  \\ \hline
812.5 &  30  &   8   &  68  & 811.11 \\ 
824.5 &  96  &  20  & 212 &  811.38 \\
822.5 &  82  &  26  & 190 & 811.47 \\
826.5 & 104 &  20  & 228 &  811.49 \\
814.5 &  40  &  16  &  96  &  811.51 \\
820.5 &  70  &  28  & 168 &  811.53 \\
816.5  &  52 &  20  & 124 &  811.58 \\
818.5  &  60 &  26  & 146 &  811.59 \\ 
828.5  & 114 & 16  & 244 &  811.66 \\ 
\hline	
\end{array}$
\end{minipage} 
\end{table}


The tables were determined as follows.
	For  each nucleus,
the smallest value of $N_0$ was determined by sequentially filling the single-particle states of a spherical harmonic oscillator.
	 They show, not surprisingly, that only for the doubly closed-shell nucleus $^{16}$O is the lowest-energy mean-field state spherical with $\lambda=\mu=0$.
	 Consistent with observations and previous analyses \cite{Morinaga56, BrownG66},  it was also not surprising to obtain, as the first excited mean-field state of $^{16}$O, the intrinsic state of a strongly deformed  rotational band with $\delta N_0 =4$ harmonic oscillator quanta more than that of the spherical  closed-shell state.
	
The table also shows that, for each nucleus considered, the state with the minimum value of $N_0$ 
has lowest mean-field energy $E_\sigma$.
	However, it must be remembered that, for a rotational nucleus, $E_\sigma$ is the mean-field energy of its deformed intrinsic state and that the energy of its zero angular-momentum component will be considerably lower.
	Thus, for example. one can expect the lower angular-momentum states of the much more strongly deformed  $N_0(\lambda,\mu) = 824.5(96,20)$ states of $^{168}$Er to
fall below those of the (812.5(30.8) states.


The tables exhibit a number of significant results.
One is that the lowest three irreps in the tables for $^{12}$C and $^{16}$O are in accord with those observed and as calculated in references \cite{DreyfussLDDB13, RoweTW06}.
	Another is that the order by increasing values of $E_\sigma$ differs markedly from the ordering by increasing values of $N_0$ as  would be obtained if the 
minimum-energy lowest-weight states were  eigenstates  of spherical harmonic-oscillator Hamiltonians.
	Also, as expected \cite{CastelRZ90}, the minimum  energies of the shape-consistent lowest-weight states obtained in this way are given almost entirely by those with
$\lambda >\mu$, consistent with the observed dominance of prolate over oblate deformations \cite{BohrM75, HamamotoM09}.

In concluding this section, it is recalled that, whereas the modification of HF applications by the inclusion of short-range pairing interactions is known to result in a partial restoration of spherical symmetry, it can similarly be expected that an extension of the AMN unified model to take account of pair coupling could result in reduced deformation and the enhancement of axial symmetry.
However, this is a possibility that remains to be explored. 

	It is also observed that among a set of Sp$(3,\Rb)$ irreps with a  common value of 
$N_0 = \sigma_1 + \sigma_2 + \sigma_3$, which are irreps having lowest-weight states of  spherical harmonic-oscillator energy $N_0 \hbar\omega_0$, the irrep of maximal space symmetry is determined to be that with a maximally deformed 
lowest-weight state.
	However, it is to be expected that the short-range interactions of the nuclear Hamiltonian, such as pairing interactions, will tend to favour axially symmetric and less deformed irreps with lower values of 
$\langle \sigma,\omega|\hat H | \sigma,\omega\rangle$, for a realistic Hamiltonian as discussed further in the concluding section of this paper.


\section{A model calculation}
\begin{figure}[ht]
\centerline{\includegraphics[width=3.5 in]{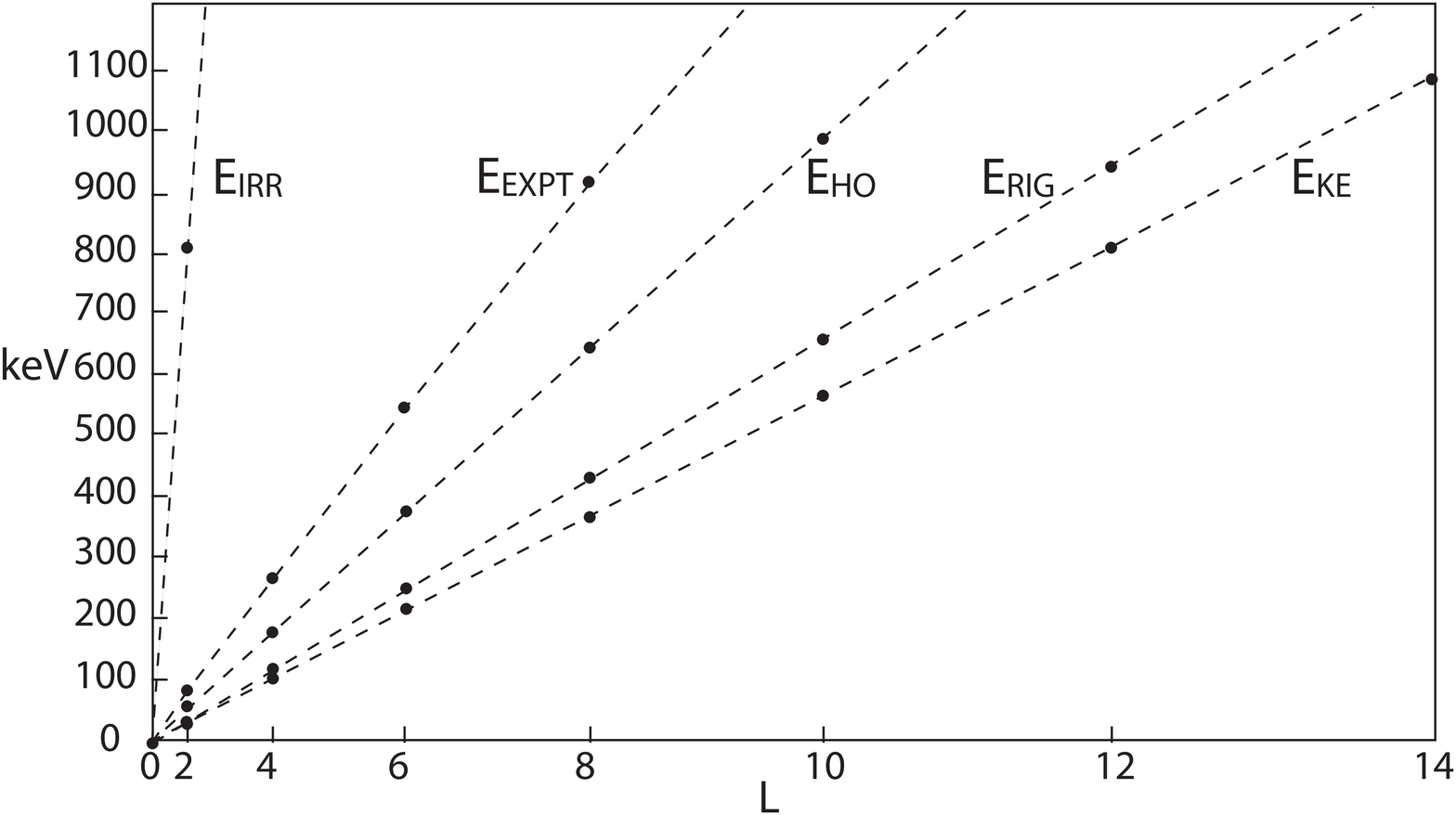}}
\caption{ \label{fig1} \footnotesize
Energy levels, labelled by ${\rm E}_{\rm EXPT}$, of the ground-state rotational band of $^{166}$Er 
and corresponding  energies for irrotational and rigid-body flows of the shape-consistent intrinsic state for the  ${\rm Sp}(3,\Rb)$ irrep 
$\langle 327\tfrac12, 249\tfrac12,249\tfrac12\rangle$.
	The kinetic energies, ${\rm E}_{\rm KE}$, were calculated for states angular-momentum projected from the shape-consistent intrinsic state as defined in section \ref{sect:V.SpST}.
	The energy levels ${\rm E}_{\rm HO}$ are those which, in addition to the kinetic energies  include the harmonic-oscillator potential energies of the angular-momentum-projected states. 
}
\end{figure}

An important property of ${\rm Sp}(3,\Rb) \times {\rm SU(2)}_S  \times {\rm SU(2)}_T$ unified-model irreps  is that they can  be explored in mean-field theory one at a time more easily and more usefully than an independent-particle irrep of a general HF calculation.
	The results, in figure \ref{fig1}, were derived for an axially symmetric ${\rm Sp}(3,\Rb) \times{\rm SU(2)}_S  \times {\rm SU(2)}_T$ irrep $\langle 327\tfrac12, 249\tfrac12,249\tfrac12\rangle$ of spin $S=0$ and minimum isospin to illustrate some of the possibilities.
	This  irrep  was used previously \cite{BahriR00}, to fit the lower-energy rotational states of  $^{166}$Er and the E2 transitions between them with a schematic interaction, without the use of an effective charge.
	According to the estimates of Jarrio et al.\,\cite{JarrioWR91},  the experimentally most appropriate irrep  would have been triaxial.
	However,  an axially symmetric representation is chosen for a first study of the model  because the results can be derived by purely algebraic methods.
	In particular, the kinetic energies of the predicted rotational states can be calculated, without resorting to numerical approximations, because the kinetic energy observable of nuclear states is an element of the Sp$(3,\Rb)$ Lie algebra and because rotational states of good angular momentum can be analytically projected from an axially symmetric intrinsic state, as shown in reference \cite{RoweBB00}.	
	The calculations  were repeated for other axially symmetric irreps and determined to be
 characteristic of any axially symmetric irrep.

	In particular, the kinetic energies of the predicted rotational states can be calculated, without resorting to numerical approximations, because the kinetic energy observable of nuclear states is an element of the Sp$(3,\Rb)$ Lie algebra and because rotational states of good angular momentum can be analytically projected from an axially symmetric intrinsic state, as shown in reference \cite{RoweBB00}.

Figure \ref{fig1} shows observed rotational energy levels of $^{166}$Er and the corresponding energies  for rigid-body and irrotational flow rotations of the intrinsic shape-consistent lowest-weight state for the $\langle 327\tfrac12, 249\tfrac12,249\tfrac12\rangle$ symplectic model irrep.
	Also shown are the kinetic energies for the angular-momentum states projected from this intrinsic state.
	An unexpected result is that they are not only less than half the observed rotational energies, they are also less than those for rigid-body rotations.  
	This implies that the observed rotational energies cannot be linear combinations of rigid and irrotational flow kinetic energies.
	The obvious interpretation is that they are sums of kinetic and potential energies.
	This is illustrated by the improvement of the results with the addition of the harmonic-oscillator potential energies  shown by ${\rm E}_{\rm HO}$ in the figure.
	It indicates that an essential  step towards understanding the dynamics of a nuclear rotational model is to calculate the separate kinetic- and potential-energy components of mean-field rotational energies.
	Thus the AMN unified model has very different implications for the dynamics of nuclear rotations to that of the standard unified model.
	Unfortunately, full AMN unified model calculations are not currently possible by purely algebraic methods for an interacting-nucleon potential energy, as they are for the kinetic 
and harmonic-oscillator potential energies of an axially symmetric irrep with
angular-momentum projection.
	It is nevertheless possible by numerical methods \cite{LeeC72} and much simpler and much more informative, than that of a standard HF calculation.

\section{Summary and conclusions}

This paper has presented a fully quantal algebraic many-nucleon version of the phenomenological Bohr-Mottelson unified model of the rotations and shape vibrations of deformed nuclei.
	Unlike the standard HF-RPA mean-field theory, which has a single minimum-energy intrinsic state, the AMN unified model has many intrinsic states and correspondingly many unified-model representations that together span the Hilbert spaces of nuclei.		
	As a result, the AMN unified model expresses the many-nucleon Hilbert space of a nucleus as a sum of optimal unified model subspaces with the property that all isoscalar E2  transition matrix elements between the states of its different subspaces are precisely zero.

	
Only the maximally deformed subset of minimum spin $S=0$ and minimum isospin $T=T_0= \tfrac12(N-Z)$ irreps has been considered in this paper.
	However, an extension to states of non-zero spin and isospin $T> \tfrac12(N-Z)$ is straightforward.
	The present focus is directed towards developing a realistic model that can be used for the analysis of experimental data,  for understanding the structures of nuclear rotational states, and for explaining the emergence of shape coexistence in nuclear physics.
	A major objective is also to develop means to interpret the results of the sophisticated program to study the many-nucleon states of deformed nuclei in the symmetry-adapted no-core shell model \cite{DytrychLDRWRBLB18}.


The results obtained in this paper take advantage of the  Nambu-Goldstone \cite{NambuJL61, GoldstoneSW62} interpretation of a broken symmetry in mean-field theory. 
	It interprets HF theory as providing a separation of the relatively slow rotational degrees of freedom of a deformed relative to those of its complementary intrinsic degrees of freedom in accordance with Born-Oppenheimer theory \cite{BornO27x}.
	It is recognised, however, that the HF constraint to independent-particle states could also be responsible for the emergence of overly deformed intrinsic states.
	The standard mechanism for obtaining spherical mean-field solutions, e.g., for singly closed-shell nuclei, 
is by an extension of HF theory to HFB (HF Bogolyubov) \cite{Bogolyubov58, Belyaev59} theory and thereby taking account of the, sometimes partial, restoration of rotational symmetry by the $J=0$ pair-coupling interactions between like nucleons.
	A more desirable solution is to retain the assumption that the rotations of a strongly deformed nucleus are adiabatic relative to its intrinsic degrees of freedom but admit a coupling between the intrinsic states of the model by $J=0$ pairing interactions without breaking nucleon number conservation.
	Hopefully, this can be achieved by the admission of intrinsic seniority correlations in coupling the intrinsic states of a rotor as expressed in the spherical shell-model seniority coupling scheme by Flowers and Edmonds \cite{Flowers52, EdmondsF52a, EdmondsF52b} and in a corresponding pair-coupling model by Kerman and others \cite{Kerman61, Lorazo70, AllaartBBSG88}.

\bibliographystyle{apsrev}
\bibliography{master}

\begin{thebibliography}{86}
\expandafter\ifx\csname natexlab\endcsname\relax\def\natexlab#1{#1}\fi
\expandafter\ifx\csname bibnamefont\endcsname\relax
  \def\bibnamefont#1{#1}\fi
\expandafter\ifx\csname bibfnamefont\endcsname\relax
  \def\bibfnamefont#1{#1}\fi
\expandafter\ifx\csname citenamefont\endcsname\relax
  \def\citenamefont#1{#1}\fi
\expandafter\ifx\csname url\endcsname\relax
  \def\url#1{\texttt{#1}}\fi
\expandafter\ifx\csname urlprefix\endcsname\relax\def\urlprefix{URL }\fi
\providecommand{\bibinfo}[2]{#2}
\providecommand{\eprint}[2][]{\url{#2}}

\bibitem[{\citenamefont{Bohr and Mottelson}(1975)}]{BohrM75}
\bibinfo{author}{\bibfnamefont{A.}~\bibnamefont{Bohr}} \bibnamefont{and}
  \bibinfo{author}{\bibfnamefont{B.~R.} \bibnamefont{Mottelson}},
  \emph{\bibinfo{title}{Nuclear Structure}}, vol.~\bibinfo{volume}{2}
  (\bibinfo{publisher}{Benjamin}, \bibinfo{address}{New York, Amsterdam},
  \bibinfo{year}{1975}), \bibinfo{note}{(republished 1998 by World Scientific,
  Singapore)}.

\bibitem[{\citenamefont{Heyde et~al.}(1983)\citenamefont{Heyde, Isacker,
  Waroquier, Wood, and Meyer}}]{HeydeIWWM83}
\bibinfo{author}{\bibfnamefont{K.~K.} \bibnamefont{Heyde}},
  \bibinfo{author}{\bibfnamefont{P.~V.} \bibnamefont{Isacker}},
  \bibinfo{author}{\bibfnamefont{M.}~\bibnamefont{Waroquier}},
  \bibinfo{author}{\bibfnamefont{J.~L.} \bibnamefont{Wood}}, \bibnamefont{and}
  \bibinfo{author}{\bibfnamefont{R.~A.} \bibnamefont{Meyer}},
  \bibinfo{journal}{Phys. Rep} \textbf{\bibinfo{volume}{102}},
  \bibinfo{pages}{291} (\bibinfo{year}{1983}).

\bibitem[{\citenamefont{Wood et~al.}(1992)\citenamefont{Wood, Heyde,
  Nazarewicz, Huyse, and {van Duppen}}}]{WoodHNHvD92}
\bibinfo{author}{\bibfnamefont{J.~L.} \bibnamefont{Wood}},
  \bibinfo{author}{\bibfnamefont{K.}~\bibnamefont{Heyde}},
  \bibinfo{author}{\bibfnamefont{W.}~\bibnamefont{Nazarewicz}},
  \bibinfo{author}{\bibfnamefont{M.}~\bibnamefont{Huyse}}, \bibnamefont{and}
  \bibinfo{author}{\bibfnamefont{P.}~\bibnamefont{{van Duppen}}},
  \bibinfo{journal}{Phys. Reports} \textbf{\bibinfo{volume}{215}},
  \bibinfo{pages}{101} (\bibinfo{year}{1992}).

\bibitem[{\citenamefont{Heyde and Wood}(2011)}]{HeydeW11}
\bibinfo{author}{\bibfnamefont{K.}~\bibnamefont{Heyde}} \bibnamefont{and}
  \bibinfo{author}{\bibfnamefont{J.~L.} \bibnamefont{Wood}},
  \bibinfo{journal}{Rev. Mod. Phys.} \textbf{\bibinfo{volume}{83}},
  \bibinfo{pages}{1467} (\bibinfo{year}{2011}).

\bibitem[{\citenamefont{{J. L. Wood and K. Heyde (Guest
  Editors)}}(2016)}]{WoodH16eds}
\bibinfo{author}{\bibnamefont{{J. L. Wood and K. Heyde (Guest Editors)}}},
  \emph{\bibinfo{title}{Shape Coexistence in Nuclei}},
  vol.~\bibinfo{volume}{43} of \emph{\bibinfo{series}{J. Phys. G. Nucl. Part.
  Phys.}} (\bibinfo{publisher}{IOP Publishing}, \bibinfo{year}{2016}).

\bibitem[{\citenamefont{Born and Oppenheimer}(1927)}]{BornO27x}
\bibinfo{author}{\bibfnamefont{M.}~\bibnamefont{Born}} \bibnamefont{and}
  \bibinfo{author}{\bibfnamefont{J.~R.} \bibnamefont{Oppenheimer}},
  \bibinfo{journal}{Annalen der Physik} \textbf{\bibinfo{volume}{389}},
  \bibinfo{pages}{457} (\bibinfo{year}{1927}), \bibinfo{note}{[translation by
  S. M. Binder in URL
  http://www.ulb.ac.be/cpm/people/scientists/bsutclif/bornop.pdf]}.

\bibitem[{\citenamefont{Bohr and Mottelson}(1953)}]{BohrM53c}
\bibinfo{author}{\bibfnamefont{A.}~\bibnamefont{Bohr}} \bibnamefont{and}
  \bibinfo{author}{\bibfnamefont{B.~R.} \bibnamefont{Mottelson}},
  \bibinfo{journal}{Dan. Mat. Fys. Medd.} \textbf{\bibinfo{volume}{{27, no.
  16}}}, \bibinfo{pages}{1} (\bibinfo{year}{1953}).

\bibitem[{\citenamefont{Bohr and Mottelson}(1955)}]{BohrM55}
\bibinfo{author}{\bibfnamefont{A.}~\bibnamefont{Bohr}} \bibnamefont{and}
  \bibinfo{author}{\bibfnamefont{B.~R.} \bibnamefont{Mottelson}},
  \bibinfo{journal}{Dan. Mat. Fys. Medd.} \textbf{\bibinfo{volume}{{30, no.
  1}}}, \bibinfo{pages}{(24pp)} (\bibinfo{year}{1955}).

\bibitem[{\citenamefont{Inglis}(1954)}]{Inglis54}
\bibinfo{author}{\bibfnamefont{D.~R.} \bibnamefont{Inglis}},
  \bibinfo{journal}{Phys. Rev.} \textbf{\bibinfo{volume}{96}},
  \bibinfo{pages}{1059} (\bibinfo{year}{1954}).

\bibitem[{\citenamefont{Inglis}(1955)}]{Inglis55}
\bibinfo{author}{\bibfnamefont{D.~R.} \bibnamefont{Inglis}},
  \bibinfo{journal}{Phys. Rev.} \textbf{\bibinfo{volume}{97}},
  \bibinfo{pages}{701} (\bibinfo{year}{1955}).

\bibitem[{\citenamefont{Thouless and Valatin}(1962)}]{ThoulessV62}
\bibinfo{author}{\bibfnamefont{D.~J.} \bibnamefont{Thouless}} \bibnamefont{and}
  \bibinfo{author}{\bibfnamefont{J.~G.} \bibnamefont{Valatin}},
  \bibinfo{journal}{Nucl. Phys.} \textbf{\bibinfo{volume}{31}},
  \bibinfo{pages}{211} (\bibinfo{year}{1962}).

\bibitem[{\citenamefont{Belyaev}(1961)}]{Belyaev61}
\bibinfo{author}{\bibfnamefont{S.~T.} \bibnamefont{Belyaev}},
  \bibinfo{journal}{Nucl. Phys.} \textbf{\bibinfo{volume}{24}},
  \bibinfo{pages}{322} (\bibinfo{year}{1961}).

\bibitem[{\citenamefont{Nathan and Nilsson}(1965)}]{NathanN65}
\bibinfo{author}{\bibfnamefont{O.}~\bibnamefont{Nathan}} \bibnamefont{and}
  \bibinfo{author}{\bibfnamefont{S.~G.} \bibnamefont{Nilsson}}, in
  \emph{\bibinfo{booktitle}{Alpha-, Beta-, and Gamma-ray Spectroscopy}}, edited
  by \bibinfo{editor}{\bibfnamefont{K.}~\bibnamefont{Siegbahn}}
  (\bibinfo{publisher}{North Holland}, \bibinfo{address}{Amsterdam},
  \bibinfo{year}{1965}), pp. \bibinfo{pages}{601--700}.

\bibitem[{\citenamefont{Rosensteel and Rowe}(1977)}]{RosensteelR77a}
\bibinfo{author}{\bibfnamefont{G.}~\bibnamefont{Rosensteel}} \bibnamefont{and}
  \bibinfo{author}{\bibfnamefont{D.~J.} \bibnamefont{Rowe}},
  \bibinfo{journal}{Phys. Rev. Lett.} \textbf{\bibinfo{volume}{38}},
  \bibinfo{pages}{10} (\bibinfo{year}{1977}).

\bibitem[{\citenamefont{Rosensteel and Rowe}(1980)}]{RosensteelR80}
\bibinfo{author}{\bibfnamefont{G.}~\bibnamefont{Rosensteel}} \bibnamefont{and}
  \bibinfo{author}{\bibfnamefont{D.~J.} \bibnamefont{Rowe}},
  \bibinfo{journal}{Ann. Phys. \textup(NY\textup)}
  \textbf{\bibinfo{volume}{126}}, \bibinfo{pages}{343} (\bibinfo{year}{1980}).

\bibitem[{\citenamefont{Draayer et~al.}(1984)\citenamefont{Draayer, Weeks, and
  Rosensteel}}]{DraayerWR84}
\bibinfo{author}{\bibfnamefont{J.~P.} \bibnamefont{Draayer}},
  \bibinfo{author}{\bibfnamefont{K.~J.} \bibnamefont{Weeks}}, \bibnamefont{and}
  \bibinfo{author}{\bibfnamefont{G.}~\bibnamefont{Rosensteel}},
  \bibinfo{journal}{Nucl. Phys. A} \textbf{\bibinfo{volume}{413}},
  \bibinfo{pages}{215} (\bibinfo{year}{1984}).

\bibitem[{\citenamefont{Rosensteel et~al.}(1894)\citenamefont{Rosensteel,
  Draayer, and Weeks}}]{RosensteelDW84}
\bibinfo{author}{\bibfnamefont{G.}~\bibnamefont{Rosensteel}},
  \bibinfo{author}{\bibfnamefont{J.~P.} \bibnamefont{Draayer}},
  \bibnamefont{and} \bibinfo{author}{\bibfnamefont{K.}~\bibnamefont{Weeks}},
  \bibinfo{journal}{Nucl. Phys. A} \textbf{\bibinfo{volume}{419}},
  \bibinfo{pages}{1} (\bibinfo{year}{1894}).

\bibitem[{\citenamefont{Park et~al.}(1984)\citenamefont{Park, Carvalho,
  Vassanji, Rowe, and Rosensteel}}]{ParkCVRR84}
\bibinfo{author}{\bibfnamefont{P.}~\bibnamefont{Park}},
  \bibinfo{author}{\bibfnamefont{J.}~\bibnamefont{Carvalho}},
  \bibinfo{author}{\bibfnamefont{M.}~\bibnamefont{Vassanji}},
  \bibinfo{author}{\bibfnamefont{D.~J.} \bibnamefont{Rowe}}, \bibnamefont{and}
  \bibinfo{author}{\bibfnamefont{G.}~\bibnamefont{Rosensteel}},
  \bibinfo{journal}{Nucl. Phys. A} \textbf{\bibinfo{volume}{414}},
  \bibinfo{pages}{93} (\bibinfo{year}{1984}).

\bibitem[{\citenamefont{Bahri and Rowe}(2000)}]{BahriR00}
\bibinfo{author}{\bibfnamefont{C.}~\bibnamefont{Bahri}} \bibnamefont{and}
  \bibinfo{author}{\bibfnamefont{D.~J.} \bibnamefont{Rowe}},
  \bibinfo{journal}{Nucl. Phys. A} \textbf{\bibinfo{volume}{662}},
  \bibinfo{pages}{125} (\bibinfo{year}{2000}).

\bibitem[{\citenamefont{Elliott}({1958})}]{Elliott58ab}
\bibinfo{author}{\bibfnamefont{J.~P.} \bibnamefont{Elliott}},
  \bibinfo{journal}{Proc. Roy. Soc. (London)} \textbf{\bibinfo{volume}{A245}},
  \bibinfo{pages}{{128, 562}} (\bibinfo{year}{{1958}}).

\bibitem[{\citenamefont{Kunz and Mosel}(1979)}]{KunzM79}
\bibinfo{author}{\bibfnamefont{J.}~\bibnamefont{Kunz}} \bibnamefont{and}
  \bibinfo{author}{\bibfnamefont{U.}~\bibnamefont{Mosel}},
  \bibinfo{journal}{Nucl. Phys. A} \textbf{\bibinfo{volume}{323}},
  \bibinfo{pages}{271} (\bibinfo{year}{1979}).

\bibitem[{\citenamefont{Radomski}(1976)}]{Radomski76}
\bibinfo{author}{\bibfnamefont{M.}~\bibnamefont{Radomski}},
  \bibinfo{journal}{Phys. Rev. C} \textbf{\bibinfo{volume}{14}},
  \bibinfo{pages}{1704} (\bibinfo{year}{1976}).

\bibitem[{\citenamefont{Fleckner et~al.}(1980)\citenamefont{Fleckner, Kunz,
  Mosel, and Wuest}}]{FlecknerKMW80}
\bibinfo{author}{\bibfnamefont{J.}~\bibnamefont{Fleckner}},
  \bibinfo{author}{\bibfnamefont{J.}~\bibnamefont{Kunz}},
  \bibinfo{author}{\bibfnamefont{U.}~\bibnamefont{Mosel}}, \bibnamefont{and}
  \bibinfo{author}{\bibfnamefont{E.}~\bibnamefont{Wuest}},
  \bibinfo{journal}{Nucl. Phys. A} \textbf{\bibinfo{volume}{339}},
  \bibinfo{pages}{227} (\bibinfo{year}{1980}).

\bibitem[{\citenamefont{de~Guerra}(1980)}]{MoyaDeGuerra80}
\bibinfo{author}{\bibfnamefont{E.~M.} \bibnamefont{de~Guerra}},
  \bibinfo{journal}{Annals of Phys.} \textbf{\bibinfo{volume}{128}},
  \bibinfo{pages}{286} (\bibinfo{year}{1980}).

\bibitem[{\citenamefont{Berdichevsky et~al.}(1988)\citenamefont{Berdichevsky,
  Sarriguren, {Moya de Guerra}, Nishimura, and Sprung}}]{BerdichevskySMdGNS88}
\bibinfo{author}{\bibfnamefont{D.}~\bibnamefont{Berdichevsky}},
  \bibinfo{author}{\bibfnamefont{P.}~\bibnamefont{Sarriguren}},
  \bibinfo{author}{\bibfnamefont{E.}~\bibnamefont{{Moya de Guerra}}},
  \bibinfo{author}{\bibfnamefont{M.}~\bibnamefont{Nishimura}},
  \bibnamefont{and} \bibinfo{author}{\bibfnamefont{D.~W.~L.}
  \bibnamefont{Sprung}}, \bibinfo{journal}{Phys. Rev. C}
  \textbf{\bibinfo{volume}{38}}, \bibinfo{pages}{338} (\bibinfo{year}{1988}).

\bibitem[{\citenamefont{Mikhailov et~al.}(1996)\citenamefont{Mikhailov,
  Briancon, and Quentin}}]{MikhailovBQ96}
\bibinfo{author}{\bibfnamefont{I.~N.} \bibnamefont{Mikhailov}},
  \bibinfo{author}{\bibfnamefont{C.}~\bibnamefont{Briancon}}, \bibnamefont{and}
  \bibinfo{author}{\bibfnamefont{P.}~\bibnamefont{Quentin}},
  \bibinfo{journal}{Physics of {P}articles and {N}uclei}
  \textbf{\bibinfo{volume}{27}}, \bibinfo{pages}{121} (\bibinfo{year}{1996}).

\bibitem[{\citenamefont{Zickendraht}(1971)}]{Zickendraht71}
\bibinfo{author}{\bibfnamefont{W.}~\bibnamefont{Zickendraht}},
  \bibinfo{journal}{J. Math. Phys.} \textbf{\bibinfo{volume}{12}},
  \bibinfo{pages}{1663} (\bibinfo{year}{1971}).

\bibitem[{\citenamefont{Filippov et~al.}(1973)\citenamefont{Filippov,
  Ovcharenko, and Steshenko}}]{FilippovOS72}
\bibinfo{author}{\bibfnamefont{G.~F.} \bibnamefont{Filippov}},
  \bibinfo{author}{\bibfnamefont{V.~I.} \bibnamefont{Ovcharenko}},
  \bibnamefont{and} \bibinfo{author}{\bibfnamefont{A.~I.}
  \bibnamefont{Steshenko}}, in \emph{\bibinfo{booktitle}{Proceedings of the
  International Symposium on Present Status and Novel Developments in the
  Nuclear Many-Body Problem: Roma, 1972}}, edited by
  \bibinfo{editor}{\bibfnamefont{F.}~\bibnamefont{Calogero}} \bibnamefont{and}
  \bibinfo{editor}{\bibfnamefont{C.}~\bibnamefont{{Ciofi degli Atti}}}
  (\bibinfo{publisher}{Editrice Compositori}, \bibinfo{address}{Bologna},
  \bibinfo{year}{1973}), pp. \bibinfo{pages}{627--668}.

\bibitem[{\citenamefont{Gulshani and Rowe}(1976)}]{GulshaniR76}
\bibinfo{author}{\bibfnamefont{P.}~\bibnamefont{Gulshani}} \bibnamefont{and}
  \bibinfo{author}{\bibfnamefont{D.~J.} \bibnamefont{Rowe}},
  \bibinfo{journal}{Can. J. Phys.} \textbf{\bibinfo{volume}{54}},
  \bibinfo{pages}{970} (\bibinfo{year}{1976}).

\bibitem[{\citenamefont{Weaver et~al.}(1976)\citenamefont{Weaver, Cusson, and
  Biedenharn}}]{WeaverCB76}
\bibinfo{author}{\bibfnamefont{L.}~\bibnamefont{Weaver}},
  \bibinfo{author}{\bibfnamefont{R.~Y.} \bibnamefont{Cusson}},
  \bibnamefont{and} \bibinfo{author}{\bibfnamefont{L.~C.}
  \bibnamefont{Biedenharn}}, \bibinfo{journal}{Ann. Phys. (N.Y.)}
  \textbf{\bibinfo{volume}{102}}, \bibinfo{pages}{493} (\bibinfo{year}{1976}).

\bibitem[{\citenamefont{Rowe and Rosensteel}(1979)}]{RoweR79}
\bibinfo{author}{\bibfnamefont{D.~J.} \bibnamefont{Rowe}} \bibnamefont{and}
  \bibinfo{author}{\bibfnamefont{G.}~\bibnamefont{Rosensteel}},
  \bibinfo{journal}{J. Math. Phys.} \textbf{\bibinfo{volume}{20}},
  \bibinfo{pages}{465} (\bibinfo{year}{1979}).

\bibitem[{\citenamefont{Rowe}(1985)}]{Rowe85}
\bibinfo{author}{\bibfnamefont{D.~J.} \bibnamefont{Rowe}},
  \bibinfo{journal}{Rep. Prog. Phys.} \textbf{\bibinfo{volume}{48}},
  \bibinfo{pages}{1419} (\bibinfo{year}{1985}).

\bibitem[{\citenamefont{Gloeckner and Lawson}(1974)}]{GloecknerL74}
\bibinfo{author}{\bibfnamefont{D.~H.} \bibnamefont{Gloeckner}}
  \bibnamefont{and} \bibinfo{author}{\bibfnamefont{R.~D.}
  \bibnamefont{Lawson}}, \bibinfo{journal}{Phys. Lett. B}
  \textbf{\bibinfo{volume}{53}}, \bibinfo{pages}{313} (\bibinfo{year}{1974}).

\bibitem[{\citenamefont{Rowe}(1984)}]{Rowe84}
\bibinfo{author}{\bibfnamefont{D.~J.} \bibnamefont{Rowe}}, \bibinfo{journal}{J.
  Math. Phys.} \textbf{\bibinfo{volume}{25}}, \bibinfo{pages}{2662}
  (\bibinfo{year}{1984}).

\bibitem[{\citenamefont{Rowe et~al.}(1984)\citenamefont{Rowe, Rosensteel, and
  Carr}}]{RoweRC84}
\bibinfo{author}{\bibfnamefont{D.~J.} \bibnamefont{Rowe}},
  \bibinfo{author}{\bibfnamefont{G.}~\bibnamefont{Rosensteel}},
  \bibnamefont{and} \bibinfo{author}{\bibfnamefont{R.}~\bibnamefont{Carr}},
  \bibinfo{journal}{J. Phys. A: Math. Gen.} \textbf{\bibinfo{volume}{17}},
  \bibinfo{pages}{L399} (\bibinfo{year}{1984}).

\bibitem[{\citenamefont{Quesne}(1986)}]{Quesne86}
\bibinfo{author}{\bibfnamefont{C.}~\bibnamefont{Quesne}}, \bibinfo{journal}{J.
  Math. Phys.} \textbf{\bibinfo{volume}{27}}, \bibinfo{pages}{428}
  (\bibinfo{year}{1986}).

\bibitem[{\citenamefont{Hecht}(1987)}]{Hecht87}
\bibinfo{author}{\bibfnamefont{K.~T.} \bibnamefont{Hecht}},
  \emph{\bibinfo{title}{The Vector Coherent State Method and its Application to
  Problems of Higher Symmetries}}, Lecture Notes in Physics
  (\bibinfo{publisher}{Springer-Verlag}, \bibinfo{address}{Berlin/Heidelberg},
  \bibinfo{year}{1987}).

\bibitem[{\citenamefont{Glauber}(1966)}]{Glauber66}
\bibinfo{author}{\bibfnamefont{R.~J.} \bibnamefont{Glauber}},
  \bibinfo{journal}{Phys. Lett.} \textbf{\bibinfo{volume}{21}},
  \bibinfo{pages}{650} (\bibinfo{year}{1966}).

\bibitem[{\citenamefont{Perelomov}(1972)}]{Perelomov72}
\bibinfo{author}{\bibfnamefont{A.~M.} \bibnamefont{Perelomov}},
  \bibinfo{journal}{Commun. Math. Phys.} \textbf{\bibinfo{volume}{26}},
  \bibinfo{pages}{222} (\bibinfo{year}{1972}).

\bibitem[{\citenamefont{Klauder and {{B.-S.} Skagerstam}}(1985)}]{KlauderB-S85}
\bibinfo{editor}{\bibfnamefont{J.~R.} \bibnamefont{Klauder}} \bibnamefont{and}
  \bibinfo{editor}{\bibnamefont{{{B.-S.} Skagerstam}}}, eds.,
  \emph{\bibinfo{title}{Coherent states: applications in physics and
  mathematical physics}} (\bibinfo{publisher}{World Scientific},
  \bibinfo{year}{1985}).

\bibitem[{\citenamefont{Kostant}(1970)}]{Kostant70}
\bibinfo{author}{\bibfnamefont{B.}~\bibnamefont{Kostant}},
  \bibinfo{journal}{Lecture Notes in Math.} \textbf{\bibinfo{volume}{170}},
  \bibinfo{pages}{87} (\bibinfo{year}{1970}).

\bibitem[{\citenamefont{Souriau}(1970)}]{Souriau70}
\bibinfo{author}{\bibfnamefont{J.~M.} \bibnamefont{Souriau}},
  \emph{\bibinfo{title}{Structure des syt{\'e}mes dynamiques}}
  (\bibinfo{publisher}{Dunod}, \bibinfo{address}{Paris}, \bibinfo{year}{1970}).

\bibitem[{\citenamefont{Rosensteel}(1986)}]{Rosensteel86}
\bibinfo{author}{\bibfnamefont{G.}~\bibnamefont{Rosensteel}},
  \bibinfo{journal}{Int. J. Theor. Phys.} \textbf{\bibinfo{volume}{25}},
  \bibinfo{pages}{553} (\bibinfo{year}{1986}).

\bibitem[{\citenamefont{Rowe and Basserman}(1976)}]{RoweB76}
\bibinfo{author}{\bibfnamefont{D.~J.} \bibnamefont{Rowe}} \bibnamefont{and}
  \bibinfo{author}{\bibfnamefont{R.}~\bibnamefont{Basserman}},
  \bibinfo{journal}{Can. J. Phys.} \textbf{\bibinfo{volume}{54}},
  \bibinfo{pages}{1941} (\bibinfo{year}{1976}).

\bibitem[{\citenamefont{Rowe and Rosensteel}(1980)}]{RoweR80}
\bibinfo{author}{\bibfnamefont{D.~J.} \bibnamefont{Rowe}} \bibnamefont{and}
  \bibinfo{author}{\bibfnamefont{G.}~\bibnamefont{Rosensteel}},
  \bibinfo{journal}{Ann. Phys. (N.Y.)} \textbf{\bibinfo{volume}{126}},
  \bibinfo{pages}{198} (\bibinfo{year}{1980}).

\bibitem[{\citenamefont{Rowe and Ryman}(1980)}]{RoweR80a}
\bibinfo{author}{\bibfnamefont{D.~J.} \bibnamefont{Rowe}} \bibnamefont{and}
  \bibinfo{author}{\bibfnamefont{A.}~\bibnamefont{Ryman}},
  \bibinfo{journal}{Phys. Rev. Lett.} \textbf{\bibinfo{volume}{45}},
  \bibinfo{pages}{401} (\bibinfo{year}{1980}).

\bibitem[{\citenamefont{Rowe and Wood}(2010)}]{RoweWood10}
\bibinfo{author}{\bibfnamefont{D.~J.} \bibnamefont{Rowe}} \bibnamefont{and}
  \bibinfo{author}{\bibfnamefont{J.~L.} \bibnamefont{Wood}},
  \emph{\bibinfo{title}{Fundamentals of Nuclear Models: Foundational Models.}}
  (\bibinfo{publisher}{World Scientific}, \bibinfo{address}{Singapore},
  \bibinfo{year}{2010}).

\bibitem[{\citenamefont{Rowe}(1966{\natexlab{a}})}]{Rowe66a}
\bibinfo{author}{\bibfnamefont{D.~J.} \bibnamefont{Rowe}},
  \bibinfo{journal}{Nucl. Phys.} \textbf{\bibinfo{volume}{80}},
  \bibinfo{pages}{209} (\bibinfo{year}{1966}{\natexlab{a}}).

\bibitem[{\citenamefont{Rowe}(1966{\natexlab{b}})}]{Rowe66b}
\bibinfo{author}{\bibfnamefont{D.~J.} \bibnamefont{Rowe}},
  \bibinfo{journal}{Nucl. Phys.} \textbf{\bibinfo{volume}{85}},
  \bibinfo{pages}{365} (\bibinfo{year}{1966}{\natexlab{b}}).

\bibitem[{\citenamefont{Bohm and Pines}(1953)}]{BohmP53}
\bibinfo{author}{\bibfnamefont{D.}~\bibnamefont{Bohm}} \bibnamefont{and}
  \bibinfo{author}{\bibfnamefont{D.}~\bibnamefont{Pines}},
  \bibinfo{journal}{Phys. Rev.} \textbf{\bibinfo{volume}{92}},
  \bibinfo{pages}{609} (\bibinfo{year}{1953}).

\bibitem[{\citenamefont{Rowe}(1968)}]{Rowe68}
\bibinfo{author}{\bibfnamefont{D.~J.} \bibnamefont{Rowe}},
  \bibinfo{journal}{Rev. Mod. Phys.} \textbf{\bibinfo{volume}{40}},
  \bibinfo{pages}{153} (\bibinfo{year}{1968}).

\bibitem[{\citenamefont{Rowe}(1970)}]{Rowebook70}
\bibinfo{author}{\bibfnamefont{D.~J.} \bibnamefont{Rowe}},
  \emph{\bibinfo{title}{Nuclear Collective Motion: Models and Theory}}
  (\bibinfo{publisher}{Methuen}, \bibinfo{address}{London},
  \bibinfo{year}{1970}), \bibinfo{note}{reprinted by World Scientific in 2010}.

\bibitem[{\citenamefont{Ring and Schuck}(1980)}]{RingS80}
\bibinfo{author}{\bibfnamefont{P.}~\bibnamefont{Ring}} \bibnamefont{and}
  \bibinfo{author}{\bibfnamefont{P.}~\bibnamefont{Schuck}},
  \emph{\bibinfo{title}{The Nuclear Many-Body Problem}}
  (\bibinfo{publisher}{Springer-Verlag}, \bibinfo{address}{New York},
  \bibinfo{year}{1980}).

\bibitem[{\citenamefont{Rowe and Basserman}(1974)}]{RoweB74}
\bibinfo{author}{\bibfnamefont{D.~J.} \bibnamefont{Rowe}} \bibnamefont{and}
  \bibinfo{author}{\bibfnamefont{R.}~\bibnamefont{Basserman}},
  \bibinfo{journal}{Nucl. Phys. A} \textbf{\bibinfo{volume}{220}},
  \bibinfo{pages}{404} (\bibinfo{year}{1974}).

\bibitem[{\citenamefont{Marumori}(1977)}]{Marumori77}
\bibinfo{author}{\bibfnamefont{T.}~\bibnamefont{Marumori}},
  \bibinfo{journal}{Prog. Theor. Phys.} \textbf{\bibinfo{volume}{57}},
  \bibinfo{pages}{112} (\bibinfo{year}{1977}).

\bibitem[{\citenamefont{Reinhard and Goeke}(1978)}]{ReinhardG78}
\bibinfo{author}{\bibfnamefont{P.~G.} \bibnamefont{Reinhard}} \bibnamefont{and}
  \bibinfo{author}{\bibfnamefont{K.}~\bibnamefont{Goeke}},
  \bibinfo{journal}{Nucl. Phys. A} \textbf{\bibinfo{volume}{312}},
  \bibinfo{pages}{121} (\bibinfo{year}{1978}).

\bibitem[{\citenamefont{Rowe and Rosensteel}(1982)}]{RoweR82}
\bibinfo{author}{\bibfnamefont{D.~J.} \bibnamefont{Rowe}} \bibnamefont{and}
  \bibinfo{author}{\bibfnamefont{G.}~\bibnamefont{Rosensteel}},
  \bibinfo{journal}{Phys. Rev. C} \textbf{\bibinfo{volume}{25}},
  \bibinfo{pages}{3236} (\bibinfo{year}{1982}).

\bibitem[{\citenamefont{Rowe}(1982)}]{Rowe82}
\bibinfo{author}{\bibfnamefont{D.~J.} \bibnamefont{Rowe}},
  \bibinfo{journal}{Nucl. Phys. A} \textbf{\bibinfo{volume}{391}},
  \bibinfo{pages}{307} (\bibinfo{year}{1982}).

\bibitem[{\citenamefont{Matsuyanagi et~al.}(2016)\citenamefont{Matsuyanagi,
  Matsuo, Nakatsukasa, Yoshida, Hinohara, and Sato}}]{MatsuyanagiMNYHS16}
\bibinfo{author}{\bibfnamefont{K.}~\bibnamefont{Matsuyanagi}},
  \bibinfo{author}{\bibfnamefont{M.}~\bibnamefont{Matsuo}},
  \bibinfo{author}{\bibfnamefont{T.}~\bibnamefont{Nakatsukasa}},
  \bibinfo{author}{\bibfnamefont{K.}~\bibnamefont{Yoshida}},
  \bibinfo{author}{\bibfnamefont{N.}~\bibnamefont{Hinohara}}, \bibnamefont{and}
  \bibinfo{author}{\bibfnamefont{K.}~\bibnamefont{Sato}}, \bibinfo{journal}{J.
  of Phys. G: Nucl. Part. Phys.} \textbf{\bibinfo{volume}{43}},
  \bibinfo{pages}{024006(20pp)} (\bibinfo{year}{2016}).

\bibitem[{\citenamefont{Rowe and Wood}(2018)}]{RoweW18}
\bibinfo{author}{\bibfnamefont{D.~J.} \bibnamefont{Rowe}} \bibnamefont{and}
  \bibinfo{author}{\bibfnamefont{J.~L.} \bibnamefont{Wood}},
  \bibinfo{journal}{J. Phys. G: Nucl. Part. Phys.}
  \textbf{\bibinfo{volume}{45}}, \bibinfo{pages}{1} (\bibinfo{year}{2018}).

\bibitem[{\citenamefont{Castel et~al.}(1990)\citenamefont{Castel, Rowe, and
  Zamick}}]{CastelRZ90}
\bibinfo{author}{\bibfnamefont{B.}~\bibnamefont{Castel}},
  \bibinfo{author}{\bibfnamefont{D.}~\bibnamefont{Rowe}}, \bibnamefont{and}
  \bibinfo{author}{\bibfnamefont{L.}~\bibnamefont{Zamick}},
  \bibinfo{journal}{Phys. Lett. B} \textbf{\bibinfo{volume}{236}},
  \bibinfo{pages}{121} (\bibinfo{year}{1990}).

\bibitem[{\citenamefont{Nambu and Jona-Lasinio}(1961)}]{NambuJL61}
\bibinfo{author}{\bibfnamefont{Y.}~\bibnamefont{Nambu}} \bibnamefont{and}
  \bibinfo{author}{\bibfnamefont{G.}~\bibnamefont{Jona-Lasinio}},
  \bibinfo{journal}{Phys. Rev.} \textbf{\bibinfo{volume}{122}},
  \bibinfo{pages}{345} (\bibinfo{year}{1961}).

\bibitem[{\citenamefont{Goldstone et~al.}(1962)\citenamefont{Goldstone, Salam,
  and Weinberg}}]{GoldstoneSW62}
\bibinfo{author}{\bibfnamefont{J.}~\bibnamefont{Goldstone}},
  \bibinfo{author}{\bibfnamefont{A.}~\bibnamefont{Salam}}, \bibnamefont{and}
  \bibinfo{author}{\bibfnamefont{S.}~\bibnamefont{Weinberg}},
  \bibinfo{journal}{Phys. Rev.} \textbf{\bibinfo{volume}{127}},
  \bibinfo{pages}{965} (\bibinfo{year}{1962}).

\bibitem[{\citenamefont{Lee and Cusson}(1972)}]{LeeC72}
\bibinfo{author}{\bibfnamefont{H.~C.} \bibnamefont{Lee}} \bibnamefont{and}
  \bibinfo{author}{\bibfnamefont{R.~Y.} \bibnamefont{Cusson}},
  \bibinfo{journal}{Phys. Lett. B} \textbf{\bibinfo{volume}{39}},
  \bibinfo{pages}{453} (\bibinfo{year}{1972}).

\bibitem[{\citenamefont{Cusson and Lee}(1973)}]{CussonL73}
\bibinfo{author}{\bibfnamefont{R.~Y.} \bibnamefont{Cusson}} \bibnamefont{and}
  \bibinfo{author}{\bibfnamefont{H.~C.} \bibnamefont{Lee}},
  \bibinfo{journal}{Nucl. Phys. A} \textbf{\bibinfo{volume}{211}},
  \bibinfo{pages}{429} (\bibinfo{year}{1973}).

\bibitem[{\citenamefont{Morinaga}(1956)}]{Morinaga56}
\bibinfo{author}{\bibfnamefont{H.}~\bibnamefont{Morinaga}},
  \bibinfo{journal}{Phys. Rev.} \textbf{\bibinfo{volume}{101}},
  \bibinfo{pages}{254} (\bibinfo{year}{1956}).

\bibitem[{\citenamefont{Brown and Green}(1966)}]{BrownG66}
\bibinfo{author}{\bibfnamefont{G.~E.} \bibnamefont{Brown}} \bibnamefont{and}
  \bibinfo{author}{\bibfnamefont{A.~M.} \bibnamefont{Green}},
  \bibinfo{journal}{Nucl. Phys.} \textbf{\bibinfo{volume}{75}},
  \bibinfo{pages}{401} (\bibinfo{year}{1966}).

\bibitem[{\citenamefont{Nilsson}(1955)}]{Nilsson55}
\bibinfo{author}{\bibfnamefont{S.~G.} \bibnamefont{Nilsson}},
  \bibinfo{journal}{Mat. Fys. Medd. Dan. Vid. Selsk.}
  \textbf{\bibinfo{volume}{29}} (\bibinfo{year}{1955}).

\bibitem[{\citenamefont{Lamm}(1969)}]{Lamm69}
\bibinfo{author}{\bibfnamefont{I.~{\relax{-}L}.} \bibnamefont{Lamm}},
  \bibinfo{journal}{Nucl. Phys. A} \textbf{\bibinfo{volume}{125}},
  \bibinfo{pages}{504} (\bibinfo{year}{1969}).

\bibitem[{\citenamefont{Bender et~al.}(2003)\citenamefont{Bender, Heenen, and
  Reinhard}}]{BenderHR03}
\bibinfo{author}{\bibfnamefont{M.}~\bibnamefont{Bender}},
  \bibinfo{author}{\bibfnamefont{P.-H.} \bibnamefont{Heenen}},
  \bibnamefont{and} \bibinfo{author}{\bibfnamefont{P.-G.}
  \bibnamefont{Reinhard}}, \bibinfo{journal}{Rev. Mod. Phys.}
  \textbf{\bibinfo{volume}{75}}, \bibinfo{pages}{121} (\bibinfo{year}{2003}).

\bibitem[{\citenamefont{Erler et~al.}(2011)\citenamefont{Erler, Kl{\"u}pfel,
  and Reinhard}}]{ErlerKR11}
\bibinfo{author}{\bibfnamefont{J.}~\bibnamefont{Erler}},
  \bibinfo{author}{\bibfnamefont{P.}~\bibnamefont{Kl{\"u}pfel}},
  \bibnamefont{and} \bibinfo{author}{\bibfnamefont{P.-G.}
  \bibnamefont{Reinhard}}, \bibinfo{journal}{Journal of Physics G: Nuclear and
  Particle Physics} \textbf{\bibinfo{volume}{38}}, \bibinfo{pages}{033101(43)}
  (\bibinfo{year}{2011}),
  \urlprefix\url{http://stacks.iop.org/0954-3899/38/i=3/a=033101}.

\bibitem[{\citenamefont{Rowe}(2013)}]{RoweRabida13}
\bibinfo{author}{\bibfnamefont{D.~J.} \bibnamefont{Rowe}},
  \bibinfo{journal}{AIP Conf. Proc.} \textbf{\bibinfo{volume}{1541}},
  \bibinfo{pages}{104} (\bibinfo{year}{2013}), \bibinfo{note}{(arXiv:1304.6115
  [nucl-th])}.

\bibitem[{\citenamefont{Dreyfuss et~al.}(2013)\citenamefont{Dreyfuss, Launey,
  Dytrych, Draayer, and Bahri}}]{DreyfussLDDB13}
\bibinfo{author}{\bibfnamefont{A.}~\bibnamefont{Dreyfuss}},
  \bibinfo{author}{\bibfnamefont{K.~D.} \bibnamefont{Launey}},
  \bibinfo{author}{\bibfnamefont{T.}~\bibnamefont{Dytrych}},
  \bibinfo{author}{\bibfnamefont{J.~P.} \bibnamefont{Draayer}},
  \bibnamefont{and} \bibinfo{author}{\bibfnamefont{C.}~\bibnamefont{Bahri}},
  \bibinfo{journal}{Phys. Lett. B} \textbf{\bibinfo{volume}{727}},
  \bibinfo{pages}{511} (\bibinfo{year}{2013}).

\bibitem[{\citenamefont{Rowe et~al.}(2006)\citenamefont{Rowe, Thiamova, and
  Wood}}]{RoweTW06}
\bibinfo{author}{\bibfnamefont{D.~J.} \bibnamefont{Rowe}},
  \bibinfo{author}{\bibfnamefont{G.}~\bibnamefont{Thiamova}}, \bibnamefont{and}
  \bibinfo{author}{\bibfnamefont{J.~L.} \bibnamefont{Wood}},
  \bibinfo{journal}{Phys. Rev. Lett.} \textbf{\bibinfo{volume}{97}},
  \bibinfo{pages}{202501} (\bibinfo{year}{2006}).

\bibitem[{\citenamefont{Hamamoto and Mottelson}(2009)}]{HamamotoM09}
\bibinfo{author}{\bibfnamefont{I.}~\bibnamefont{Hamamoto}} \bibnamefont{and}
  \bibinfo{author}{\bibfnamefont{B.~R.} \bibnamefont{Mottelson}},
  \bibinfo{journal}{Phys. Rev. C} \textbf{\bibinfo{volume}{79}},
  \bibinfo{pages}{034317} (\bibinfo{year}{2009}).

\bibitem[{\citenamefont{Jarrio et~al.}(1991)\citenamefont{Jarrio, Wood, and
  Rowe}}]{JarrioWR91}
\bibinfo{author}{\bibfnamefont{M.}~\bibnamefont{Jarrio}},
  \bibinfo{author}{\bibfnamefont{J.~L.} \bibnamefont{Wood}}, \bibnamefont{and}
  \bibinfo{author}{\bibfnamefont{D.~J.} \bibnamefont{Rowe}},
  \bibinfo{journal}{Nucl. Phys. A} \textbf{\bibinfo{volume}{528}},
  \bibinfo{pages}{409} (\bibinfo{year}{1991}).

\bibitem[{\citenamefont{Rowe et~al.}(2000)\citenamefont{Rowe, Bartlett, and
  Bahri}}]{RoweBB00}
\bibinfo{author}{\bibfnamefont{D.~J.} \bibnamefont{Rowe}},
  \bibinfo{author}{\bibfnamefont{S.}~\bibnamefont{Bartlett}}, \bibnamefont{and}
  \bibinfo{author}{\bibfnamefont{C.}~\bibnamefont{Bahri}},
  \bibinfo{journal}{Phys. Lett. B} \textbf{\bibinfo{volume}{472}},
  \bibinfo{pages}{227} (\bibinfo{year}{2000}).

\bibitem[{\citenamefont{Dytrych et~al.}(2018)\citenamefont{Dytrych, Launey, and
  {Draayer et al.}}}]{DytrychLDRWRBLB18}
\bibinfo{author}{\bibfnamefont{T.}~\bibnamefont{Dytrych}},
  \bibinfo{author}{\bibfnamefont{K.~D.} \bibnamefont{Launey}},
  \bibnamefont{and} \bibinfo{author}{\bibfnamefont{J.~P.} \bibnamefont{{Draayer
  et al.}}}, \bibinfo{journal}{arXiv:1810.05757v1 [nucl-th]}
  (\bibinfo{year}{2018}).

\bibitem[{\citenamefont{Bogolyubov}(1958)}]{Bogolyubov58}
\bibinfo{author}{\bibfnamefont{N.~N.} \bibnamefont{Bogolyubov}},
  \bibinfo{journal}{Nuovo Cim.} \textbf{\bibinfo{volume}{7}},
  \bibinfo{pages}{794} (\bibinfo{year}{1958}).

\bibitem[{\citenamefont{Belyaev}(1959)}]{Belyaev59}
\bibinfo{author}{\bibfnamefont{S.~T.} \bibnamefont{Belyaev}},
  \bibinfo{journal}{Mat. Fys. Medd. Dan. Vid. Selsk}
  \textbf{\bibinfo{volume}{{31, no. 11}}}, \bibinfo{pages}{(55pp)}
  (\bibinfo{year}{1959}).

\bibitem[{\citenamefont{Flowers}(1952)}]{Flowers52}
\bibinfo{author}{\bibfnamefont{B.~H.} \bibnamefont{Flowers}},
  \bibinfo{journal}{Proc. Roy. Soc. London} \textbf{\bibinfo{volume}{A212}},
  \bibinfo{pages}{248} (\bibinfo{year}{1952}).

\bibitem[{\citenamefont{Edmonds and Flowers}(1952{\natexlab{a}})}]{EdmondsF52a}
\bibinfo{author}{\bibfnamefont{A.~R.} \bibnamefont{Edmonds}} \bibnamefont{and}
  \bibinfo{author}{\bibfnamefont{B.~H.} \bibnamefont{Flowers}},
  \bibinfo{journal}{Proc. Roy. Soc. (London) A} \textbf{\bibinfo{volume}{214}},
  \bibinfo{pages}{515} (\bibinfo{year}{1952}{\natexlab{a}}).

\bibitem[{\citenamefont{Edmonds and Flowers}(1952{\natexlab{b}})}]{EdmondsF52b}
\bibinfo{author}{\bibfnamefont{A.~R.} \bibnamefont{Edmonds}} \bibnamefont{and}
  \bibinfo{author}{\bibfnamefont{B.~H.} \bibnamefont{Flowers}},
  \bibinfo{journal}{Proc. Roy. Soc. (London) A} \textbf{\bibinfo{volume}{215}},
  \bibinfo{pages}{120} (\bibinfo{year}{1952}{\natexlab{b}}).

\bibitem[{\citenamefont{Kerman}(1961)}]{Kerman61}
\bibinfo{author}{\bibfnamefont{A.~K.} \bibnamefont{Kerman}},
  \bibinfo{journal}{Ann. Phys. \textup(NY\textup)}
  \textbf{\bibinfo{volume}{12}}, \bibinfo{pages}{300} (\bibinfo{year}{1961}).

\bibitem[{\citenamefont{Lorazo}(1970)}]{Lorazo70}
\bibinfo{author}{\bibfnamefont{B.}~\bibnamefont{Lorazo}},
  \bibinfo{journal}{Nucl. Phys.} \textbf{\bibinfo{volume}{153}},
  \bibinfo{pages}{255} (\bibinfo{year}{1970}).

\bibitem[{\citenamefont{Allaart et~al.}(1988)\citenamefont{Allaart, Bonsignori,
  Savoia, and Gambhir}}]{AllaartBBSG88}
\bibinfo{author}{\bibfnamefont{K.}~\bibnamefont{Allaart}},
  \bibinfo{author}{\bibfnamefont{G.}~\bibnamefont{Bonsignori}},
  \bibinfo{author}{\bibfnamefont{M.}~\bibnamefont{Savoia}}, \bibnamefont{and}
  \bibinfo{author}{\bibfnamefont{T.~K.} \bibnamefont{Gambhir}},
  \bibinfo{journal}{Phys. Reports} \textbf{\bibinfo{volume}{169}},
  \bibinfo{pages}{209} (\bibinfo{year}{1988}).

\end{thebibliography}

\end{document}